\documentclass[useAMS,usenatbib]{mn2e}
\usepackage[]{natbib,refname,amsmath,amssymb,times}
\bibpunct{(}{)}{;}{a}{}{,}

\usepackage[utf8]{inputenc}
\usepackage{eurosym,graphicx,hyperref,epsfig,latexsym}

\usepackage{color}

\renewcommand{\d}{\mathrm{d}}

\newcommand{\ii}{\mathrm{i}}
\newcommand{\bea}{\begin{eqnarray}}
\newcommand{\eea}{\end{eqnarray}}
\newcommand{\be}{\begin{equation}}
\newcommand{\ee}{\end{equation}}
\newcommand{\rund}[1]{\left(#1\right)}
\newcommand{\eck}[1]{\left[ #1 \right]}

\newcommand{\vc}[1]{\mbox{\boldmath $#1$}}

\def\elabel#1{\label{eq:#1}}

\title[A new measurement in weak lensing]
{Measurement of differential magnification}
\author[Xer]%
{
Xinzhong Er$^1$\thanks{E-mail: xer@nao.cas.cn}
\\
$^1$National Astronomical Observatories, Chinese Academy of Sciences,
20A Datun Road, Beijing 100012, China
}%
\date{Accepted ---; received ---; in original form \today}
\pubyear{2014}

\begin{document}

\maketitle
\begin{abstract}
In gravitational lensing, the magnification effect changes the
luminosity and size of a background galaxy. If the image sizes are not
small compared to the scale over which the magnification and shear
vary, higher-order distortions occur which are termed differential
magnification.  We give an approximation of the magnification gradient
for several halo models. Assuming a symmetric distribution of source
brightness, estimates for the differential magnification are obtained
and then tested with simulations. One of the main uncertainties of our
estimators comes from the finite resolution of the image. We study the
strength of our method with the resolution of current and future
telescopes. We point out that out method is a potential approach to estimate
the first flexion, and can be used to study galaxy and cluster mass
profiles.
\end{abstract}

\section{Introduction}

Gravitational lensing is a powerful tool in modern cosmology
\citep[see e.g.][]{1992grle.book.....S}. It provides a method to
estimate mass distributions
without assuming the properties or the kinematics of the matter, and
has been widely used on different scales, e.g. from galaxies and galaxy
clusters to large scale structures. For instance, the strong lensing
arc statistics are sensitive to the cosmological parameters
\citep[e.g.][]{2005ApJ...635..795L}, image time delays can be used to
measure the Hubble constant \citep[e.g.][]{2011A&A...536A..53C,
  2013MNRAS.432..679C}.  Moreover, mass reconstructions using weak
lensing image distortions are widely applied in studying galaxies and
clusters \citep[e.g.][]{2012ApJ...758..128C,2014arXiv1404.1375U}.
Flexion as a higher order weak lensing distortion, has higher
signal-to-noise ratios in the region between the typical weak lensing
and strong lensing regions \citep{2005ApJ...619..741G,bacon2006}. It is a
potential tool to estimate halo ellipticity
\citep{2011A&A...528A..52E,Er&bartelmann2012} and substructures
\citep{2010MNRAS.409..389B}. However, due to the extreme difficulty in
measuring flexion
\citep[e.g.][]{2012MNRAS.419.2215V,2013MNRAS.435..822R}, few
useable results have been achieved.

Lensing is also a powerful cosmic telescope for studying high redshift
objects \citep[e.g.][]{2012ApJ...745..155H, 2014ApJ...782L..36S,
  2013arXiv1310.6695B}. Due to the large distances of high redshift
objects, it is difficult to detect them. The lensing magnification
effect enhances the luminosity of the background objects, and
increases the probability of finding high redshift objects. However
differential magnification causes bias in the parameter estimation for
extended lensed objects, since the magnification varies with the
position within the source \citep{2013ApJ...770..110E}. In particular for
extended sources close to caustics on the source plane, the
magnification will be significantly different over the image. Thus,
there is a brightness gradient over the lensed images for sources
with symmetric surface brightness, which it is possible to detect.

The flux ratios between multiple images in strong lensing systems
provide interesting information on the lens. In particular, we can use
that information to constrain local substructures
\citep[e.g.][]{1998MNRAS.295..587M, 2005MNRAS.364.1459C,
  2005astro.ph..9252S,2010ApJ...715..793G}.
Extending this approach, we propose a simple method to measure the
magnification gradient effect by considering flux ratios between
different parts of a lensed image. The validity of our method is based
on the assumption that the image of the source galaxy is centrally
symmetric in brightness, which is roughly true for most elliptical
galaxies. For spiral or irregular galaxies, large sample statistics
have to be adopted. For the very strongly lensed sources, e.g., giant
arcs, the number of observed samples is too small to perform a
statistical study. In the very weak lensing regime, the magnification
approaches unit, and the differential effect thus can be
neglected. Therefore, the optimal region to measure the effect is at
intermediate distances, which is the moderate lensing region
\citep{2012MNRAS.422.2808M}. It is not surprising to see that this is
the same region for measuring the flexion signal, since flexion also
measures the gradient effect. In fact, the magnification gradient is
proportional to the first flexion.  Therefore, from a brightness
gradient we can estimate the gradient of surface mass density.

The basic theory is given in section 2. We discuss the properties of
some dark matter halo models in section 3, and perform a numerical
simulation in section 4. A differential magnification estimator is
given and the difficulties in using real observations are discussed.
The cosmological parameters that we use are $\Omega_{\Lambda}=0.6825$,
$\Omega_m=0.3175$, Hubble constant
$H_0=100h$km\,s$^{-1}$\,Mpc$^{-1}$ and $h=0.671$, which are based on
the results from the PLANCK project \citep{2013arXiv1303.5076P}.

\section{Basic formalism}
The basic formalism of gravitational lensing can be found in
\citet{2001PhR...340..291B}. We follow the complex
notation system for its elegance and brevity. The thin-lens approximation
is adopted, implying that the lensing mass distribution can be
projected onto the lens plane perpendicular to the line-of-sight. We
introduce angular coordinate $\vc\theta$ on the lens plane (with
$\theta$ the absolute value). The magnification due to lensing is
written as
\be
\mu(\theta) ={1\over \eck{1-\kappa(\theta)}^2 -\gamma \gamma^*(\theta)},
\ee
where $\kappa$ is convergence and $\gamma=\gamma_1+\ii\gamma_2$ is
lensing shear. The magnification gradient is thus
\bea
\delta \mu(\theta) &=& \nabla_c \rund{1 \over (1-\kappa)^2 -\gamma\gamma^*}\\
&=& \mu^2\, \eck{2(1-\kappa) {\cal F} + \gamma {\cal F}^* + \gamma^* {\cal G}},
\elabel{maggradient}
\eea
where $\nabla_c$ is the complex differential operator, and ${\cal
  F}=\nabla_c\kappa$, ${\cal G}=\nabla_c\gamma$ are the first and
second flexions \citep{schneider&er08}.  The gradient of magnification
is proportional to $\mu^2$ and is related to both shear and flexion.

We consider the orders of magnitude of the variations in
magnification. If we assume that the lensed image is in weak lensing
region, i.e. the shear is small, the product terms involving shear and
flexion can be neglected. A simple estimate of the magnification
gradient is obtained from (\ref{eq:maggradient})
\be
\delta \mu \approx 2 \mu^2 {\cal F}.
\elabel{diffmag}
\ee
It has the same dimension as flexion, $\propto \theta^{-1}$. The
differential magnification, i.e., the magnification difference for
given spatial separation $\vc\delta\theta$ can be calculated as
$\Delta\mu\equiv \vc\delta\theta\cdot \vc\delta\mu$. For a general
magnitude of flexion \citep{2007ApJ...660.1003G,2008ApJ...680....1O},
${\cal F}\sim0.03$ arcsec$^{-1}$, and for a typical scale of background
source size $\sim2$ arcsec, the variation of magnification over the
image can be significant, $|\Delta\mu/\mu|\sim10\%$.

In weak lensing, the shape of images is unchanged if the surface mass
density $\kappa$ is transformed as $\kappa\to\kappa' = \lambda \kappa
+ (1-\lambda)$, which is known as the mass-sheet degeneracy
\citep{1988ApJ...327..693G}. The real observable quantity in weak
lensing is the reduced shear $g=\gamma/(1-\kappa)$. The reduced
magnification thus becomes $\mu'^{-1}=1-gg^*$. The gradient of reduced
magnification does not have the correspond low order term as that in
(\ref{eq:diffmag})
\be
\delta\mu' = \mu'^2 (g G_1^*+g^* G_3),
\ee
where $G_1$ and $G_3$ are the reduced flexions. Since the low order
approximation term disappears in the reduced gradient, we will use
(\ref{eq:diffmag}) to calculate magnification gradients in the rest of
this paper.

\begin{figure}
\centerline{\includegraphics[width=10.0cm]{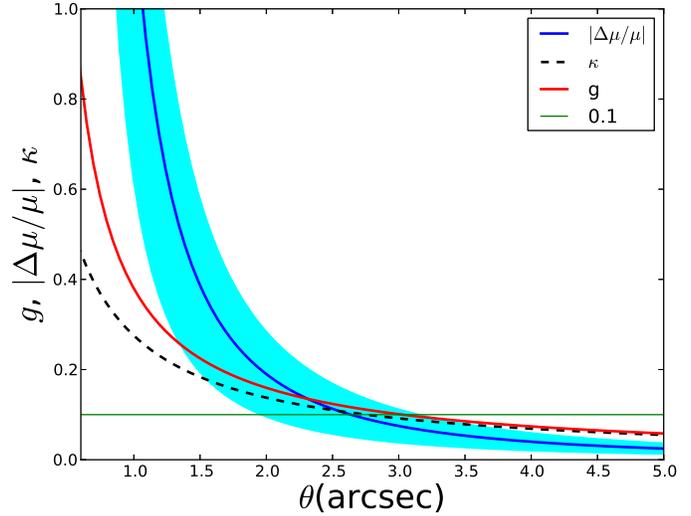}}
\caption{ The radial profile of differential magnification, reduced
  shear and convergence for an SIS halo lens. The shadow is the region
  bounded by the image size ($0.5-1.5$ arcsec). Constant value of $0.1$
  is plotted for better comparison.}
\label{fig:shear-mu}
\end{figure}

In Fig.~\ref{fig:shear-mu}, we illustrate the reduced shear
\citep{1995A&A...294..411S}, convergence and differential
magnification. We use a Singular Isothermal Sphere (SIS) lens model with
Einstein radius $\theta_E=0.55$ arcsec. The redshifts
of the lens and source are $0.1$ and $1.0$ respectively. The differential
magnification is normalized using the total magnification in order to show
the relative change over the image, and also for better
visibility. The blue line represents the result for the background image
with size of $\delta\theta=1$ arcsec. The blue shadow is the region
bounded by image size $0.5 \le \delta \theta \le 1.5$ arcsec.
We can see
that the differential magnification effect is significant within six
times the Einstein radius. This is also the expected high signal-to-noise
region for measuring flexion. At larger central distances, sample volume
and background image size are critical conditions for obtaining a usable
signal.

\section{Circular halo profiles}
In this section, we will present the predictions from differential
magnification for a variety of different lens models. A circularly
symmetric lens halo profile is adopted for simplicity, and is a good
assumption statistically \citep{2012A&A...545A..71V}. It is valid for
galaxy-galaxy lensing with a circularly averaged mean lens.

\begin{figure*}
\centerline{\includegraphics[width=8.0cm]{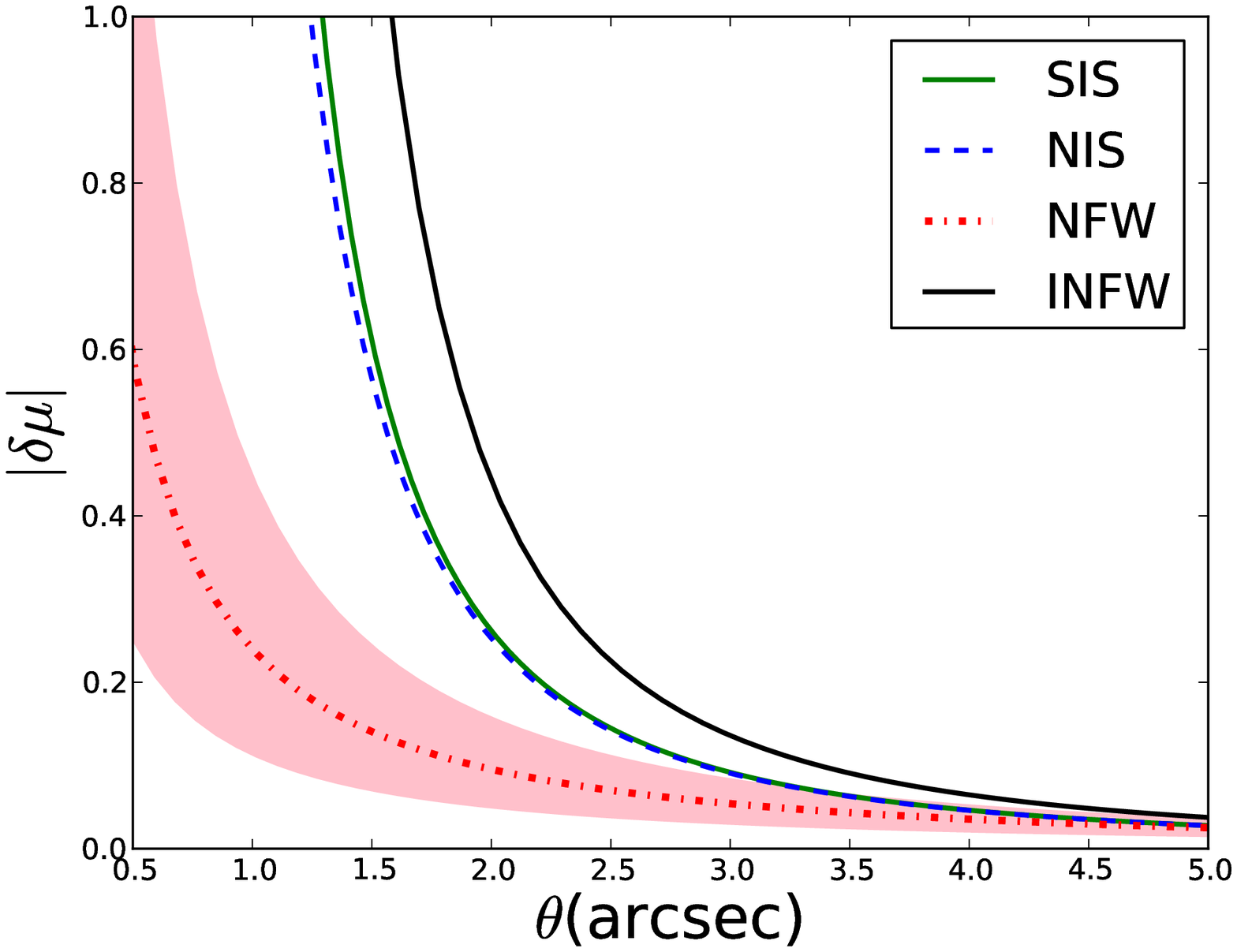}
\includegraphics[width=8.0cm]{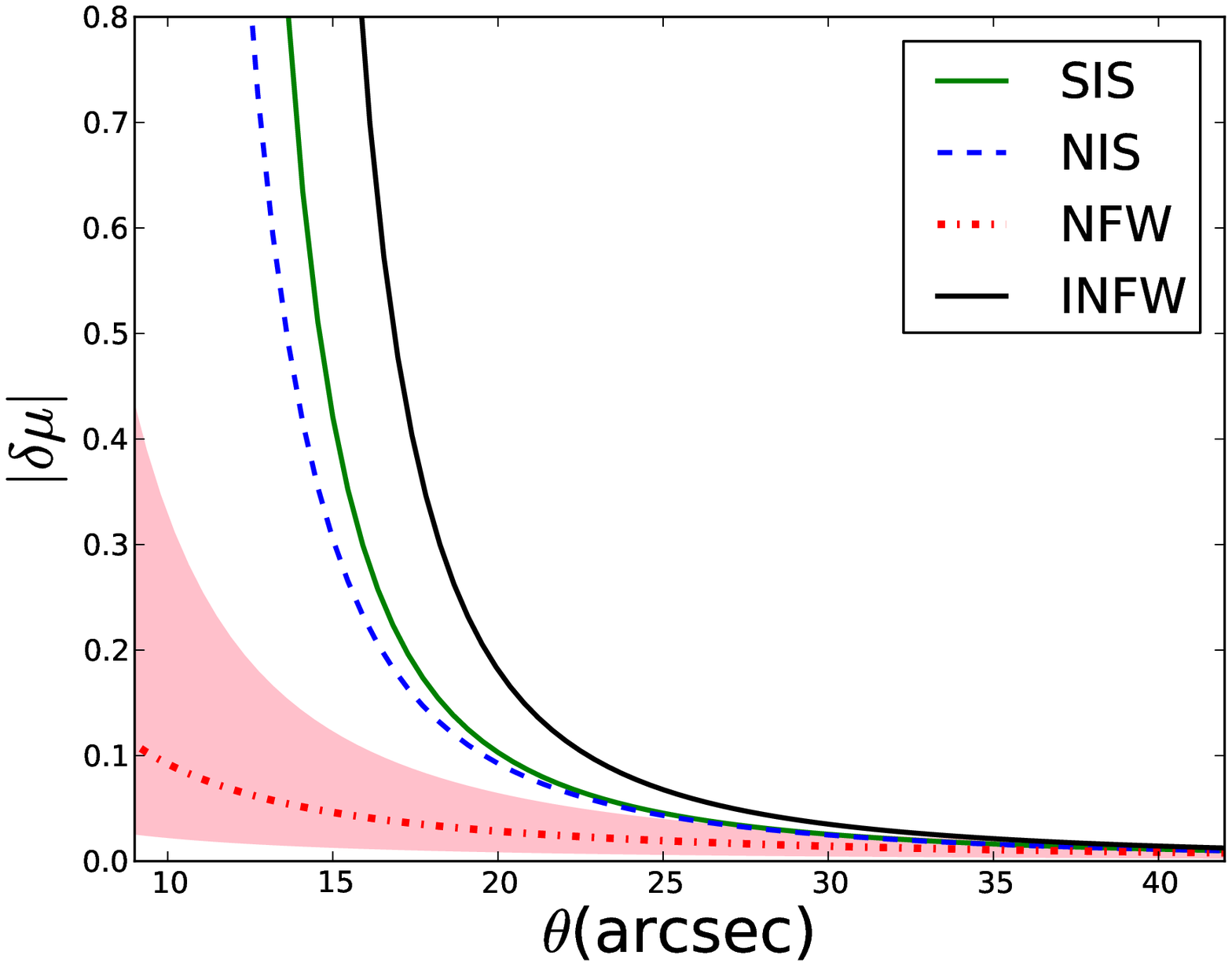}}
\caption{The magnification gradient for different halo profiles: green
  solid-SIS, dashed blue-NIS, red dotted-NFW and black solid-INFW.
  The pink shadow presents the variance due to changing of
  concentration parameter for the results of NFW halo. The left(right)
  panel is for a galaxy(cluster) sized halo. For SIS and NIS halo, we use
$\theta_E=0.55(10)$ arcsec for left(right) panel.}
\label{fig:sis2nfw}
\end{figure*}

We start from the basic form of halo model, a power-law profile.
For the projected mass density with form $\kappa=A y^{-\alpha}$, where
$y=\theta/\theta_E$ is the dimensionless central distance,
the deflection potential is
\be
\psi =
\begin{cases}
A({\rm ln} y )^2 +C_1 {\rm ln}y +C_2,\quad (\alpha=2)\\
\\
\dfrac{2 A y^{2-\alpha}}{(2-\alpha)^2} + C_1 {\rm ln}y +C_2 \quad (0<\alpha<2).
\end{cases}
\ee
Power-law profiles with a large index ($\alpha\geqslant2$) will not
be considered in the following part of this work, since it is usually
the profile at the outer range of galaxy/cluster halos, and the lensing
signal for our proposes is weak. The magnification and gradient for a
power-law halo profile are
\bea
\mu &\propto& {1 \over 1- 2 y^{-\alpha}}\qquad \\
\vc\delta \mu &\propto& {-2 \alpha y^{-\alpha-1} \over (1- 2 y^{-\alpha})^2}\, \hat\theta,
\eea
where $\hat\theta$ is the unit vector.
The widely used halo model for galaxy lens is the SIS halo ($\alpha=1$,
$\kappa=\theta_E/(2\theta)$). The magnification is
\be
\mu={\theta \over \theta -\theta_E},
\ee
and the gradient
\be
\vc\delta \mu = -{\theta_E \over (\theta - \theta_E)^2} {\rm e}^{-\ii \phi},
\ee
where $\phi$ is the position angle around the lens. To remove the
divergence of mass density as for $\theta\rightarrow 0$, one simple
modification is to cut off the distribution at small distances as
follows
\be
\kappa = {\theta_E \over 2 \sqrt{\theta^2+ \theta_c^2} },
\ee
where $\theta_c$ is a core radius within which the surface mass
density flattens off to a value $\theta_E/(2\theta_c)$. This model is known as a
Non-singular Isothermal Sphere (NIS), and its magnification and
gradient are
\bea
\mu &=& \eck{1-\dfrac{\theta}{\Theta} +
\dfrac{\theta_E^2\theta_c}{\Theta(\Theta +\theta_c)^2}}^{-1},\\
\vc\delta\mu &=& \mu^2 \eck{\theta\theta_E \over \Theta^3}{\rm e}^{\ii\phi},
\eea
where $\Theta=\sqrt{\theta^2+\theta_c^2}$.  The properties of NIS halo
behave like the SIS for $\theta\gg\theta_c$
(Fig.~\ref{fig:sis2nfw}). In the region where the differential
magnification is interesting, an SIS halo is a good approximation for
the NIS halo.

Using N-body simulations, Navarro, Frenk and White (NFW) have
shown that the equilibrium density profile of dark matter haloes can
be fitted by a universal profile\citep{1995MNRAS.275..720N,
1997ApJ...490..493N}. The NFW
halo is parameterised by mass ($m_{200}$) and concentration ($c$). The
magnification of the NFW profile is given by
\bea
\mu^{-1}= 4k_s^2\{1 - {2(1-f_k) \over x^2-1}
+ {4(f_k+{\rm ln}(x/2))(1-f_k) \over x^2(x^2-1)} \nonumber\\
 - {4(f_k+{\rm ln}(x/2))^2 \over x^4}\},
\elabel{munfw}
\eea
with function
\be
f_k=\begin{cases}
\dfrac{{\rm arcsech}x}{\sqrt{1-x^2}} \qquad(x<1);\\
\\
1 \qquad\qquad\quad (x=1);\\
\\
\dfrac{{\rm arcsec}x}{\sqrt{x^2-1}} \qquad(x>1),
\end{cases}
\ee
where $x=\theta D_d/r_s$ and $D_d$ is the angular diameter distance of
the lens. The lensing properties of the NFW halo are controlled by the
parameters $k_s$ and $r_s$. More detail about the lensing properties of the NFW
halo can be found in \citet{1996A&A...313..697B}. The magnification
gradient can be calculated from (\ref{eq:diffmag} and \ref{eq:munfw}),
and two examples are shown in Fig.~\ref{fig:sis2nfw}.

We compare the magnification gradient for a typical
galaxy-/cluster-sized halo with different profiles. For both galaxy
and cluster halos, the lens and source redshifts are $z_d=0.1$ and
$z_s=1.0$ respectively. In the left panel of Fig.~\ref{fig:sis2nfw},
the parameter of the SIS halo is the same as that in
Fig.~\ref{fig:shear-mu}, $\theta_E=0.55$ arcsec.  $\theta_c=0.2$
arcsec is used for the NIS halo. For the NFW halo, we use
$m_{200}=10^{12}\, M_{\odot}$ and vary the concentration between $7$
and $14$ (pink shaded region in left panel of Fig.~\ref{fig:sis2nfw}).
The isothermal-like NFW (INFW) halo is a mass profile more concentrated in
the centre of the halo than that of the NFW halo. The mass density in
the inner region is similar to the SIS halo, $\rho\propto
1/(r^2(1+r))$ \citep{er2013infw}. The magnification gradient for
isothermal-like NFW halo is also shown for comparison. The same
$m_{200}$ is used and the concentration parameter is $c_I=1.9$. In the
right panel, we show the properties of cluster-sized halos. For the
SIS and NIS halos, we use $\theta_E=10$ arcsec and $\theta_c=2$
arcsec. The parameters for NFW and INFW halos are:
$m_{200}=2\times10^{14}\, M_{\odot}$, $c=6$ (the shadow represents the
variation due to concentration $4-9$) and $c_I=0.8$.
In Fig.~\ref{fig:sis2nfw}, different lines show the results for
different profiles. We can see that the SIS, NIS and INFW profiles
have significant magnification gradients in our region of interesting for
galaxy(cluster) $1.5-6$($1.5-3$) times the Einstein radius. The NFW
profile has a lower gradient than the other profiles. We will confront
different difficulties in galaxy- or cluster-lens. For the
galaxy-sized halo, the image of background galaxy will be contaminated
by the lens galaxy if the separation is too small. The signal becomes
too weak to be detected with large separation. Thus in order to study
the galaxy halo properties we need two basic conditions: large lensed image;
and a galaxy-galaxy pair sample with optimal separations. For
the cluster halo, although the signal drops faster with radius (only
to $3$ times Einstein radius), the lensed image has less probability
to be contaminated by foreground galaxies. The cluster properties
are more complicated, e.g. irregular shape of halo, massive satellite
galaxies etc. Moreover, more than one critical curve can be generated
by a cluster. The magnification gradient does not monotonically decrease
with radius. The signal will not be isotropic vectors. The direction
of differential magnification is not easy to determine. In this work,
we mainly focus on the galaxy-sized halo.

\section{A ratio estimator and Numerical tests}
\begin{figure}
\centerline{\includegraphics[width=7.5cm,height=6.8cm]{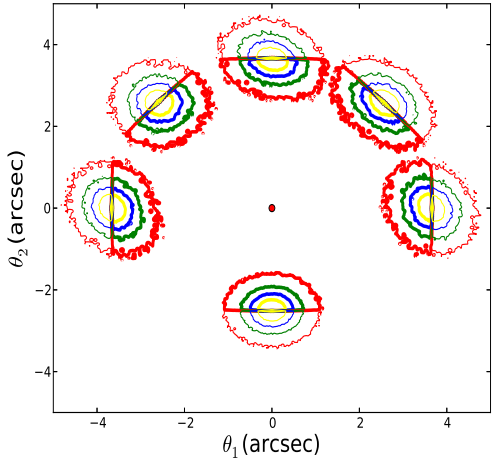}}
\caption{ The lensed image of a galaxy at different positions. The
  lens is at the origin of the figure (solid dot). An unlensed image
  is also shown at the bottom for comparison. The different color
  lines show the iso-surface brightness contours. All the images are
  separated into two parts along the centre. The thick contour lines
  show the inner part of the images. The same lens and source
  properties as in Fig.~\ref{fig:lratio} are used. Poisson noise is
  added to each pixel.}
\label{fig:fluxmap}
\end{figure}

The magnification effect cannot be measured directly, since we do not
know the intrinsic luminosity of the source galaxy. Under the
assumption that the background galaxy is uniformly bright, the gradient
of the lensed galaxy brightness represents the magnification
gradient. However, the brightness of galaxies is not uniform in
reality. Due to the finite spatial resolution, we can only estimate
the gradient on average to a limited resolution. Therefore, we use a
flux ratio to estimate the gradient effect. Basically, the lensed
image will be separated into two parts: one is close to the lens, and
the other one is far from the lens (Fig.~\ref{fig:fluxmap}).

We consider an image of a source by lensing, and denote the brightness
distribution of the source (lensed image) by $I^s(\beta)$
($I(\theta)=I^s(\beta(\theta))$, where $\beta$ is the source
coordinate). We separate the image into two parts: with subscript of
$_{\rm in}$ and $_{\rm out}$. The flux of each part will be
\bea
S^s_{\rm in,out} &=& \int_{\rm in,out}\d^2\,\beta\, I^s(\beta) =
\int_{\rm in,out} \d^2\theta\,{\rm det}A(\theta)\,I(\theta);\\
S_{\rm in,out} &=& \int_{\rm in,out}\d^2\, \theta\, I(\theta),
\eea
where $A$ is the Jacobian matrix of the lens equation, and ${\rm det}A=1/\mu$.
Thus approximately we have
\be
S_{\rm in,out}= \dfrac{\int_{\rm in,out} \d^2\theta\,\mu(\theta)}
{\int_{\rm in,out} \d^2 \theta\, }\; S_{\rm in,out}^s.
\ee
We choose the two parts of the image with equal area, i.e. $\int_{\rm
  in}\d^2\theta=\int_{\rm out}\d^2\theta$. Under the assumption that
the source is symmetric in brightness distribution ($S^s_{\rm in} =
S^s_{\rm out}$), we define a magnification ratio over the lensed
image, which can be measured by the brightness ratio
\be
\Delta\mu \equiv {\int_{\rm in}\, \d^2\theta\,\mu(\theta) \over \int_{\rm out}\,
\d^2\theta\,\mu(\theta)} = {S_{\rm in} \over S_{\rm out}}.
\elabel{lumratio}
\ee
We can see that it does not require the knowledge of intrinsic
brightness of the background galaxies. Moreover, we do not have to
assume a uniform brightness. The only assumption we need is that the
galaxy is centrally symmetrical ($I(\vc\beta)=I(-\vc\beta)$). Most
of the elliptical galaxies satisfy the condition approximately.
In general, the sizes of background galaxies in weak lensing
surveys do not exceed two or three arcsecs. The magnification ratio can
be approximated by first order correction. We thus calculate the
magnification ratio
\be
\Delta\mu = {\mu_0 - \delta\theta\delta\mu \over \mu_0 }
\approx {\mu_0 - 2 \mu_0^2 {\cal F} \delta\theta \over \mu_0 },
\elabel{delmu}
\ee
where $\delta\theta$ is the scale of the lensed galaxy, $\mu_0$ is the
mean magnification of the outer part of the lensed image. We use
(\ref{eq:diffmag}) in the approximate calculation.

\begin{figure}
\centerline{\includegraphics[width=6.0cm]{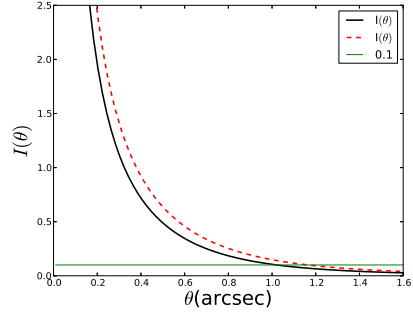}}
\caption{ The surface brightness profile of the source galaxy
  (Eq.\ref{eq:sourceb}). }
\label{fig:sourceb}
\end{figure}

We construct some numerical simulations to test the behavior of the
estimators (\ref{eq:lumratio}) and (\ref{eq:delmu}). The mock data is
generated by ray-tracing simulations. We use an SIS dark matter halo
with the same parameters as in Fig.~\ref{fig:shear-mu}:
$\theta_E=0.55$ arcsec, $z_d=0.1$, and vary the source redshifts
between $z_s=0.4$ and $z_s=3.0$. The brightness profile of the source is
given by a combination of a bulge and a disk:
\be
I(r) = I_e\, {\rm exp}\rund{-7.67\eck{(r/r_e)^{1/4}-1}}
+ I_d\, {\rm exp}(-r/r_d),
\elabel{sourceb}
\ee
where the parameters we use are: $I_e=0.3$ is the surface brightness
at the bulge effective radius $r_e=0.5$ arcsec, $I_d=1.0$ is the
central surface brightness, and $r_d=0.3$ arcsec is the scale length
of the exponential disk (the unit of the brightness is arbitrary,
since we only need the brightness ratio). The source brightness
profile is shown by solid line in Fig.~\ref{fig:sourceb}. In general,
the images of galaxies are elliptical \citep{2013MNRAS.431..477J}. We
thus use $r=\sqrt{x^2f+y^2/f}$, where $f=2/3$ is the axis ratio of the
source galaxy.

\begin{figure}
\centerline{\includegraphics[width=10.0cm]{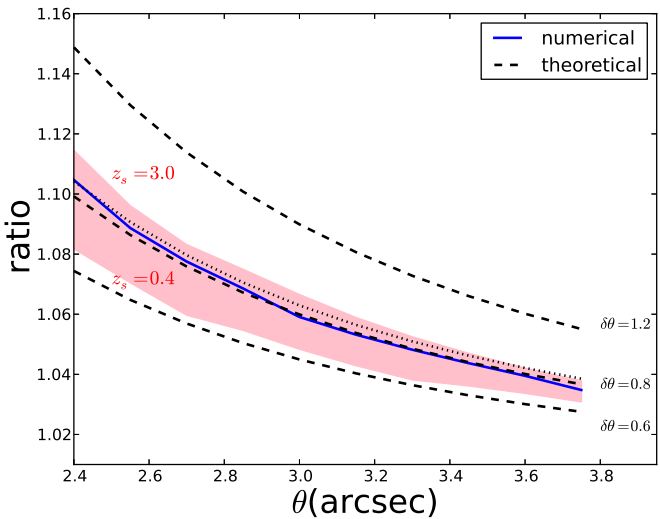}}
\caption{ The luminosity ratio of lensed image between the inner part
  and the outer part. The dashed lines are theoretical predictions for
  different image sizes; the solid line is the numerical result for a
  background source at $z_s=1.0$. The pink shadow region represents
  the range due to source redshifts ($z_s\subset [0.4,3.0]$).}
\label{fig:lratio}
\end{figure}

For the lensed image, we select $10$ pixels with maximum luminosity.
The centre of the image is identified as luminosity weighted centre of
the $10$ points. Along the centre of the image, we separate the image
into two regions (close to the lens and far from the lens), and
calculate the brightness ratio between the two regions
(Fig.~\ref{fig:fluxmap}). The results are shown in
Fig.~\ref{fig:lratio}. We show the results roughly between $4$ and $6$
times Einstein radius, which is the region to measure the weak lensing
signal. The image of the background galaxy is not contaminated by the
lens, and we still have a signal which it is possible to measure. We
study the region with weak signal in order to test the abilities of
our estimator.
The solid line represents the result from numerical simulations for
the sources at redshift $z_s=1.0$. The shadow shows the variation due
to changing of the source redshift ($[0.4:3.0]$). The dashed lines are
theoretical predictions calculated from (\ref{eq:delmu}). The three
dashed lines are the results for $\delta\theta=0.6,0.8,1.2$ arcsec
respectively. We can see that the one for $\delta\theta=0.8$ arcsec
agrees with the numerical result. The dependence on the redshift is
weaker than that on the image size.

\subsection{Size, Pixelation and other systematics}

There are several aspects may cause difficulties in the ratio measurement
using real observations, such as the brightness profile of the source
galaxy or the Point Spread Function (PSF). We will mainly discuss about
the effects due to image size and pixelation.

The size of the lensed image is ill-defined, especially for
strongly lensed images, e.g. arcs. This is also the reason that we do not
propose to perform our measurement within $1.5$ times Einstein radius.
For weakly lensed images, we can use the second order brightness moment to
estimate the size of image
\be
Q_0=\dfrac{\int \d\theta^2\, \theta \theta^* I(\theta)}
{\int \d\theta^2\, I(\theta)}.
\ee
The differential magnification scale is calculated by
$\delta\theta=C\sqrt{Q_0}/2$. The pre-factor $C$ takes account of the
shape of the lensed image. For regularly lensed elliptical images, the
differential scale along the edge is shorter than that of the centre. We use
a value of $2/3$ in this work. The orientation of the image will cause a
slightly different $\delta\theta$ (one can see the difference of the
orientation parallel or perpendicular to the direction of
magnification gradient). In Fig.~\ref{fig:lratio}, we use the mean
scale over all the image with different orientation
($\delta\theta\approx 0.8$ arcsec).
At small radii, since the magnification is slightly larger, the
size of the lensed image is also slightly larger. Therefore, the
theoretical prediction with constant image size goes below the
numerical results, and the difference becomes significant as the
radius becomes smaller. In the case of Fig.~\ref{fig:lratio}, the
magnification at $\theta=2.4$ arcsec is $1.1$ times larger than that
at $\theta=3.6$ arcsec. It enlarge the image size to
$\delta\theta=0.84$ arcsec. The prediction for that is shown by the
dotted line in Fig.~\ref{fig:lratio}, which agrees with the numerical
results at small radii.

One major problem is the finite spatial resolution. Since our
estimator is using the statistics of image brightness, high spatial
resolution is preferred. For the pixelated
image, the image centre can be easily mis-identified, and cause
systematics in measuring the brightness ratio. The method we used
($10$ points averaging) to find image centre can reduce the
fluctuations in the centre identification. For small size images
(a few hundreds pixels), the uncertainty is still significantly large.
Moreover, for low number pixel images, the estimated size of the image
has large uncertainty.

We perform numerical simulations to test the uncertainty arising due
to the pixel effect. The same lens model as in the previous section
(Fig.~\ref{fig:lratio}) is adopted. While two source brightness models
are used: one is the same as that in previous section, and the other
is $10\%$ larger than the first
($I_e=0.33,r_e=0.55,I_d=1.1,r_d=0.33$). The brightness profiles of the
two source models are given in Fig.~\ref{fig:sourceb}. Three spatial
resolutions are used in our tests. The scale of each pixel is:
$10,\,30,\,65$mas respectively.  $30,65$mas are the pixel size of the
Hubble Space Telescope in the CLASH survey \citep[see][for more
  detail]{2012ApJS..199...25P}, and $10$mas is the best resolution of
the Keck telescope using Adaptive Optics. We also add Poisson noise in
each pixel.

\begin{figure*}
\centerline{\includegraphics[width=8.5cm]{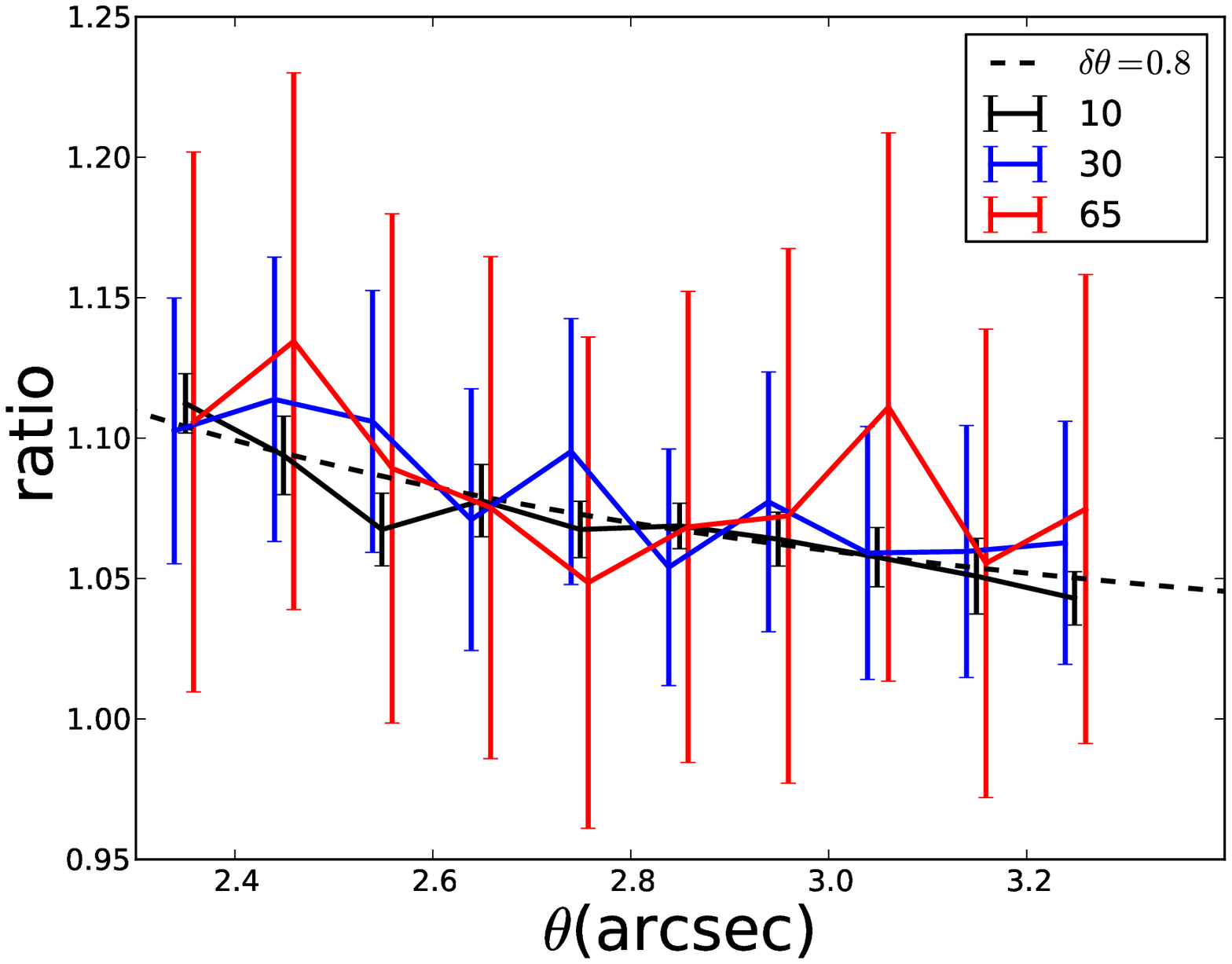}
\includegraphics[width=8.5cm]{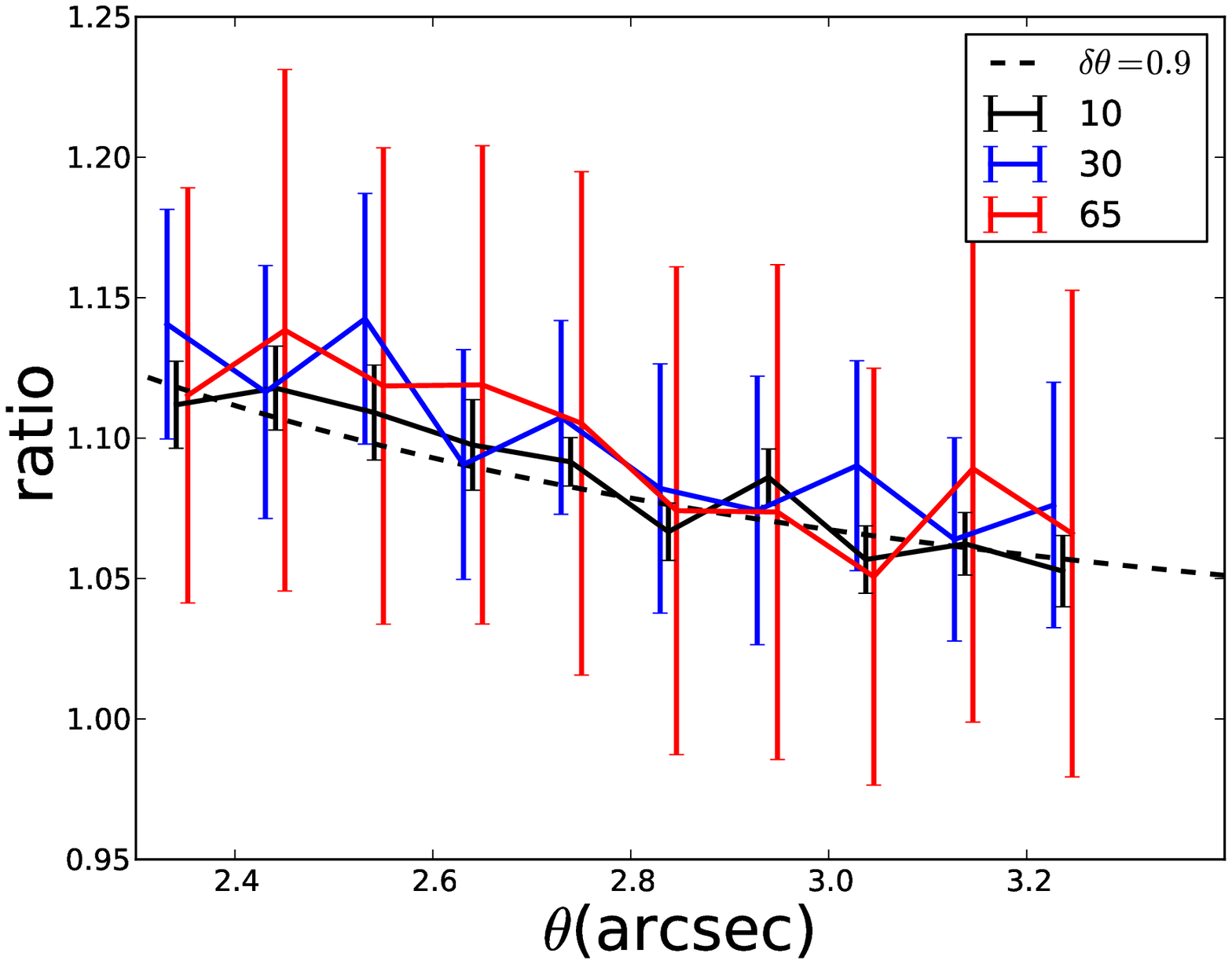}}
\caption{The luminosity ratio due to pixel effect. The left panel use
  the same lensing and source model as in Fig.~\ref{fig:lratio}, the
  right panel use larger source image. The dashed line presents the
  theoretical prediction. The red, blue and black lines represent the
  results with pixel size of $65,30,10$mas respectively. }
\label{fig:pixel}
\end{figure*}

The result is shown in Fig.~\ref{fig:pixel}. The left panel is for the
small source, and right panel is for the larger one. We apply a simple
statistic to calculate the value of each point. At each position, we
generate a set of realizations to calculate the brightness ratio. The
mean value and standard deviation over
each set of realizations are our estimated value and error.
The number of realizations for resolutions $10, 30, 65$mas are $10, 50,
100$ for the small image and $10, 30, 50$ for the large image respectively.
We can see that the results from small pixel or large image give
stable estimates with small error bars. For the large pixel, we need a
large sample size to reduce the uncertainty. The minimum required
number for given pixel size may depends on several aspects of the
observation, e.g. size of image. A simple experience from our
numerical test: the images with more than $1000$ pixels can provide
stable results using our method.

\begin{figure}
\centerline{\includegraphics[width=8.5cm]{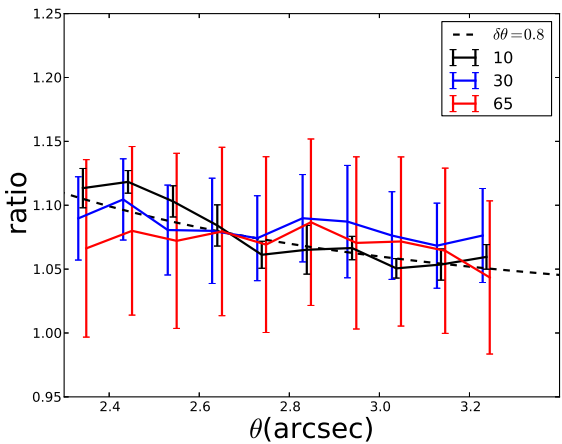}}
\caption{The luminosity ratio : the same conditions are used as left panel in
  Fig.~\ref{fig:pixel} except including a PSF. }
\label{fig:psf}
\end{figure}

Moreover, we adopt an ideal point spread function (PSF) in the
numerical tests: a circular symmetric Gaussian shape with width
$\sigma=0.05$ arcsec. The same conditions of lens, source and pixel
size as that in previous paragraph are used. In Fig.~\ref{fig:psf},
one can see that in this case, the PSF effect is not significant. Only
for $65$mas resolution, the result at small radius is depressed. This
is however strongly underestimating the systematics due to the PSF. In
reality, the PSF is not regular or symmetric, and it is not uniform
over the field of view.

Another approximation in our method is that we simply cut the lensed
image along the tangential direction to the lens. These two parts are
not exactly the two equal parts in the source plane. There is a slight
difference. This is however not significant in the region where we
are interested.

In additional tests, we also study some other aspects of source
galaxies: brightness profile, ellipticity and orientation. The
ellipticity and the orientation have a small impact to our estimator. The
brightness profile can slightly influence the result, especially
the low resolution ones. All the aspects can cause large uncertainty
when the number of pixels is small.

%

\section{Summary and discussion}

In this paper, we have studied the magnification gradient effect in
weak gravitational lensing. The gradient can be used to study the mass
profile of dark matter halos for galaxies and clusters.
We compare the signals from different halo mass models. The isothermal
sphere halo can generate higher gradients than the NFW model. The
galaxy lens is more efficient than a cluster one: up to $6$ times the
Einstein radius, the galaxy signal is strong enough to be detected
($>0.05$); while for cluster the signal drops below $0.05$ at $3$
times the Einstein radius. The difficulty for measuring galaxy-galaxy
lensing is that the background image may be contaminated by the lens
galaxy. Therefore, a sufficient sample of large image separations is
necessary. For galaxy clusters, the foreground luminous contamination
will not be a problem since the image separation is sufficiently
large. However, the galaxy cluster is more complicated than the galaxy
in mass distribution. For instance, the mass centre is ill-defined and
difficult to identify.  There is often a large offset between the true
mass centre and the bright centre galaxy
\citep[e.g.][]{2009MNRAS.398.1698S,2011MNRAS.410..417S}.  Moreover,
other aspects can also cause systematics, such as the irregular shape,
massive substructures and multiple critical curves. One needs an
accurate mass model to determine the direction of the magnification
gradient in order to study the lensed images.

One of the important applications of magnification gradients is that
they are proportional to the first flexion, and thus provide a
potential tool to estimate the first flexion. However, it is an
indirect estimation which also depends on the magnification to the
power of $2$. A tiny uncertainty in magnification can cause a
significant large error in flexion estimation. However, in the study
of dark matter ellipticity, the aspect we are interested in is the
direction of the first flexion vector. The magnitude of the flexion
which is influenced by the magnification does not effect the result,
although a high magnitude flexion will certainly provide high
signal-to-noise.

We derive the relationships between the brightness ratio of source and
image. Under the condition that the brightness of source galaxy is
symmetric, the brightness ratio provides an estimator of differential
magnification. We perform numerical experiments to study the
behaviors of the estimator. The estimator agrees with the
theoretical prediction. Due to the finite image pixel size, there are
fluctuations of our estimate. The uncertainty can be reduced by
applying large sample statistics or by using large size images.

Similar to shear measurements, the brightness method presented
here must be modified in several ways to be applicable to real
data. First, brightness can be weighted in order not to be dominated
by the very noisy outer regions of the image. However, whether it will
introduce extra bias needs careful study. Secondly, one needs to
account for the effects of the PSF. Some approaches, such as image fitting
may provide stable estimates, since the brightness profile and PSF can
be easily taken into account. Unlike shear or flexion measurement,
our estimator uses the brightness of the image. Thus, the requirement
on the image quality, i.e. PSF, may not be as high as that for shear or
flexion, but we do need small pixel size or a large image. Some more
advanced methods should be developed to reduce the uncertainty in
brightness ratio, e.g. the centre and size for images with a low
number of pixels.

Moreover, using gravitational lensing as a cosmic telescope to study
high redshift galaxies becomes more and more promising. The bias due
to differential magnification thus is important and should be taken
into account when calculating the physical parameters of lensed
objects. For high redshift lensed objects, the magnification is
usually larger than $10$. The differential magnification can easily
reach $\sim10\%$ for even extremely small images ($\sim0.1$ arcsec). Our
method (Eq.\ref{eq:diffmag}) only provides an approximated
correction for this kind of bias. For highly magnified extended
images, one has to consider the higher order corrections.

For galaxy-sized lens, we can also perform stacking of galaxy-galaxy
lensing systems. A deep survey with a large volume sample is
favored. Statistically, the effects due to asymmetry and different
brightness profiles can be reduced. However, we have to confront again
the difficulty of identifying the centre of lensed images, and how to
align the image centre. Fine pixel size is a critical issue for our
method. For low pixel number images, interpolation or fitting for the
image may provide higher accuracy. From our simulation, current
surveys, such as the Hubble Space Telescope and the Keck telescope
have sufficiently small pixel sizes to perform our
study. Future telescopes such as the Thirty Meter Telescope will almost
certainly allow very accurate estimates of the mass gradient of galaxy-
and cluster-sized lenses.

\section*{Acknowledgments}
I would like to thank the referee for suggestions and comments, and thank
Shude Mao, Jiren Liu, Keiichi Umetsu, Matthias Bartelmann, Richard J.
Long and Junqiang Ge for discussions and comments on the draft of this
paper. X.E. is supported by NSFC grant No. 11203029.

\bibliographystyle{mn2e}
\bibliography{../../../bib/lens,../../../bib/flexion,../../../bib/refcos,../../../bib/shape,../../../bib/galaxy,../../../bib/stronglens,../../../bib/refbooks}

\end{document}